\documentstyle[12pt,epsf,epsfig,pstricks,wrapfig]{article}
\begin{document}
\title{Wave-particle duality}
{\large\noindent  \bf Wave-particle duality and the zittrbewegung} \vspace{5mm}

\noindent Paul O'Hara
\footnote {\small\it
Istituto Universitario Sophia,
Via San Vito, 28 - Loppiano,
50064 Figline e Incisa Valdarno (FI)
\hspace{5mm}
email: paul.ohara@sophiauniversity.org}
%
\newtheorem{thm}{Theorem}
\newtheorem{cor}{Corollary}
\newtheorem{defn}{Definition}
\newtheorem{lem}{Lemma}

\begin{abstract}
In previous work, the Hamilton-Jacobi equation has been associated with the metrics of general relativity and shown to be a generalized Dirac equation for quantum mechanics. This lends itself to a natural definition of wave-particle duality in quantum mechanics. This theory is now further developed to show that a free spinless quantum particle moving with velocity $v$ obeys the standard wave equation of electro-magnetism. We also discuss the implications for the zitterbewegung problem and its relationship to isotropy. Moreover, it is also shown that for the theory to be consistent, the momentum defined by the Hamilton-Jacobi function presupposes the existence of a universal parameter internal to the system. In the case of particles with mass this invariant can be defined by $d\lambda=dt/m(t)$ where $t$ has the units of time and $m=m(t)$ has the units of mass. 

\noindent KEY Words: Hamilton-Jacobi equation,  invariant parameter, zitterbewugung
\end{abstract}
\makeatletter
%

\section{Introduction}
The expression wave-particle duality is associated with quantum mechanics and quantum field theory. Historically it dates back to Einstein's publication in 1905  of the photo-electric effect, in which the transmission of light previously assumed to be a wave phenomena also exhibited particle properties. In contrast, De Broglie waves associated with objects which had traditionally been considered  particles were also found to exhibit wave-like characteristics. However, it should be noted that by definition of wave-particle duality (see below) there are no particles in a classical sense. When we refer to an electron as a particle, in reality we are referring to the ``particle properties'' associated with the center-of-mass.

More formally, we treat the subject
from the perspective of general relativity. We find that there is a one-to-one correspondence between the metrics of general relativity and wave equations, which will permit us to
associate the particle momentum with its corresponding quantum mechanical
operator. This approach will allow us to explain the zitterbewegung problem in a natural manner. Before entering into the specifics, we first give a brief  overview
of the relationship between differential geometry and Clifford Algebras.

\section{Metrics and the Dirac equation\label{sec:Metrics-Dirac-eq}}

What is the role of spacetime metrics in the description of fundamental
forces and interactions? General relativity allows us to identify mass with spacetime
curvature or more precisely to identify the stress-energy-momentum tensor with the Ricci curvature tensor. 
In this theory, a planet moves along a straight line (geodesic)
in its locally flat spacetime, which corresponds to a trajectory
of a closed ellipsoid loop in the globally curved spacetime. One may also calculate approximately 
the same planetary trajectory via Newton's gravitational force over a flat spacetime or via Einstein's spacetime gravitational field equations, nevertheless they are different from both an ontological and physics perspective.
The new insights gained through the concepts and equations of general relativity involve a paradigm shift.
Unlike Newtonian mechanics, general
relativity gives a unified method to calculate the motion of planets
and the gravitational deflection of electromagnetic waves. 

\subsection{Dual equations}

We begin with an intuitive and non-rigorous approach to our methodology
by indicating two ways in which quantum mechanical wave equations
can be obtained from the metrics of general relativity, without any
explicit recourse to Lagrangians or Hamiltonians. We will then combine
the results of the two approaches into a mathematical theorem. In
this section, we will impose more rigorous constraints which
will enable us to identify the spinor formulation given here with
the usual Hilbert space formulation of quantum mechanics. In general, Einsteinian
notation will be used for index summations throughout this chapter, although at times it will 
be necessary to distinguish the time component from the space ones in which case we will do so using the $\sum$ notation.

The General Theory of Relativity associates gravitational
and electromagnetic fields with four-dimensional differential
manifolds. Consequently, the language of differential geometry is
at the heart of the subject. We denote a manifold and its local metric
tensor as $\left({\mathcal M}, g\right)$. Particles move on curves,
which in turn can be defined in terms of metrics $ds^{2}=g_{\mu\nu}dx^{\mu}dx^{\nu}$
where $(dx^{\mu})$ represents a tangent vector along the curve, expressed
in terms of differentials defined with respect to a basis. At each
point along the curve we can also assign a vector field $(\partial/\partial x^{\mu})$,
whose operation on a function $\psi$ can be associated with a gradient
$\frac{\partial\psi}{\partial x^{\mu}}$ defined over the manifold.
In effect, there is a canonical 1-1 correspondence between vectors
defined on the tangent space associated with the directional derivative
along a curve, and those which can be identified as gradients defined
over the dual space associated with the same curve: 
\begin{eqnarray}
T_{0}^{1}(M) & \leftrightarrow & T_{1}^{0}(M)\\
v^{\mu}\partial_{\mu} & \leftrightarrow & \tilde{v}_{\mu}dx^{\mu}\label{DE:gr-eq-2-1}
\end{eqnarray}

Since a manifold is a locally flat surface, we can erect an orthogonal
tetrad (vierbien) at each point. In terms of general relativity and
the Principle of Equivalence, this means that there exists a
(local) tetrad at every point of the field, such  that for
each $v\in T_{0}^{1}$ and the corresponding $\tilde{v}\in T_{1}^{0}$
we can write respectively 
\[
v=v^{a}\partial_{a}\qquad{\rm and}\qquad\tilde{v}=v_{a}dx^{a}\ .
\]
The use of tetrads is essential when we work with spinors.

We shall now exploit this 1-to-1 correspondence to identify the quantum-mechanical
wave equations as the dual of the Dirac ``square-root'' of the metric
(see also \cite{ohara-wp} and \cite{poh1-wp}). Let
\begin{eqnarray}
ds^{2}=g_{\mu\nu}dy^{\mu}dy^{\nu}=\eta_{ab}dx^{a}dx^{b}\label{eq:gr-eq-2-1}
\end{eqnarray}
where $a$ and $b$ refer to local tetrad coordinates and $\eta$
to a rigid (local) Minkowski metric. We used $(dy^{\mu})$ and $(dx^{a})$
symbols for the respective tangent vectors in order to emphasize that
these are two different coordinate systems. Associated with this metric
$\eta$ are the scalar $ds$ and a matrix $\tilde{ds}\equiv\gamma_{a}dx^{a}$
such that $\tilde{ds}^{2}=ds^{2}$. Here $\gamma_{a}$ is a Clifford
basis, which transforms as a covariant vector under coordinate transformations.
Geometrically, $\gamma_{a}dx^{a}$ is a vector and $ds$ is its norm.
There are many possible choices for the matrix
representation of $\gamma_{a}$, and the most frequently used ones
are representations of the Pauli spin matrices. Regardless of our choice 
of basis, it is important to define\
 $\{\gamma_{a}\gamma_{b}\}\equiv\gamma_{a}\gamma_{b}+\gamma_{b}\gamma_{a}=2\eta_{ab}$ in accordance with the commutation rules of a Clifford basis.

Note also that the relationship between the two metric tensors is given by $g_{\mu\nu}(x)=\eta_{ab}e_{\mu}^{a}(x)e_{\nu}^{b}(x)$
with $e_{\mu}^{a}(x)$ forming local tetrads at $x$.  Furthermore,
$ds$ can be considered as an ``eigenvalue'' of the linear operator
$\tilde{ds}$ by forming the spinor eigenvector equation $\tilde{ds}\xi=ds\xi$.

Based on the above definitions, we associate the metric 
\begin{eqnarray}
ds^{2}=g_{\mu\nu}dy^{\mu}dy^{\nu}=\eta_{ab}dx^{a}dx^{b}\label{eq:gr-eq-3-1}
\end{eqnarray}
with the spinor equation: 
\begin{eqnarray}
ds\xi=\gamma_{a}dx^{a}\xi\label{eq:gr-eq-4-1}
\end{eqnarray}
This  equation, in a natural way, associates spinors
directly with the metrics of general relativity. Moreover, in agreement
with the general theory of eigenvectors, if $\xi$ is a solution so
also is $\psi(z_{0},z_{1},z_{2},z_{3})\xi$ where $\psi$ is any complex
scalar valued function. Indeed, there is no reason why $\psi$ cannot
be an $L^{2}$ function that corresponds to a quantum-mechanical wave
function. It is also worth keeping in mind that every operator has
its corresponding eigenfuctions and by definition, these eigenfuctions
are invariant with respect to the action of the operator. Although in
quantum mechanics the act of measurement interferes with the initial
state of the wave function, nevertheless each eigenvector corresponds to an invariant
state with respect to the action of the operator. Usually this is
framed in the language of the projection postulate and in this context
quantum theorists talk about a collapsed wave function. However, this
is a misnomer. In reality, it is not that the wave function has collapsed
but rather it has been transformed into another one, with the eigenvector
still remaining invariant. For example, if we consider the rotation
of the earth, the eigenvector will correspond to its axis of rotation
while all other points are constantly in motion because of the rotation.

Just as each vector $\frac{\partial}{\partial x^{a}}$ can be mapped to a
dual one-form $dx^{a}$, similarily, the $\tilde{ds}$ matrix
above can be seen as the dual of the expression 
$\tilde{\partial}_{s}\equiv\gamma^{a}\frac{\partial}{\partial x^{a}}$,
where $\gamma^{a}$ is defined by the relationship $\{\gamma^{a},\gamma_{b}\}=2\delta_{b}^{a}$, or equivalently $\eta_{ab}\gamma^{a}=\gamma_{b}$.

If we let $s$ describe the length of a particle's trajectory along
a curve $(x^{0}(s),x^{1}(s),x^{2}(s),x^{3}(s))\in\left({\mathcal M},g\right)$
then $s$ can be regarded as an independent parameter with an associated
1-form $ds$, which is the dual of the tangent vector $\partial_{s}$.
Note that in terms of the basis vectors for $T_{p}\left(\mathcal{M}\right)$
and $T_{p}^{*}\left(\mathcal{M}\right)$ we can write $\partial_{s}=\frac{\partial x^{a}}{\partial s}\partial_{a}$
and $ds=\frac{\partial s}{\partial x^{a}}dx^{a}$. It also follows
that this dual map is given by $ds\partial_{s}\equiv\frac{\partial s}{\partial x^{i}}\frac{\partial x^{i}}{\partial s}=1$.
Putting these two results together allows us to consider equation
\ref{eq:gr-eq-4-1} as the dual of the equation: 
\begin{eqnarray}
\frac{\partial\psi}{\partial s}=\gamma^{a}\frac{\partial\psi}{\partial x^{a}},\label{eq:gr-eq-6-1}
\end{eqnarray}
where $\frac{\partial}{\partial s}$ refers to differentiation along
a curve parametrized by $s$. Geometrically, equation (\ref{eq:gr-eq-6-1})
is the dual of equation (\ref{eq:gr-eq-4-1}).

We shall refer to equation (\ref{eq:gr-eq-6-1}) as a (generalized)
Dirac equation and later show how it relates to the usual form of
this equation. At times, we shall also loosely refer to it as a ``dual
wave-equation''. \textbf{This 1-to-1 map between equations (\ref{eq:gr-eq-4-1})
and (\ref{eq:gr-eq-6-1}) formally defines what we mean by wave-particle
duality.} They both go together and one cannot exist without the other.
The metric corresponds to the particle property and the wave equation
to its wave property. Given the metric, we can write down the wave
and conversely, given the wave we can write down the metric. It is
only a matter of convention whether we begin first with the wave and
then the metric or vice-versa. It remains to better understand the
origin and properties of this wave from a physical point of view. The first step in this understanding is to show 
the relationship between Minkowski space, null metrics and the classical wave equation. This is formulated more precisely
 in the following theorem and lemmas.
\begin{thm}
\label{thm:DE-1 } Let $ds^{2}=g_{\mu\nu}dy^{\mu}dy^{\nu}=\eta_{ab}dx^{a}dx^{b}$ define a metric along 
a curve defined over $T^0_1$ then
$$ ds^{2}=\eta_{ab}dx^{a}dx^{b}\iff 0=v^2dt^2-dx^2_1-dx^2_2-dx^2_3$$ where
$\  v_i=\frac{dx_i}{dt}, \ v=|v_i|\ .$
\end{thm}
\noindent{\textbf Proof:} Note $ds^{2}=c^{2}d\tau^{2}$  from which it follows that 
\begin{eqnarray}
c^{2}d\tau^{2} & = & c^{2}dt^{2}-\sum^3_{i=1}dx^{2}_i=c^{2}dt^{2}\left(1-\frac{v^{2}}{c^{2}}\right)\\
{\rm Therefore} \qquad 0 & = & c^{2}dt^{2}-c^{2}d\tau^{2}-\sum_idx^{2}_i\\
\iff\qquad 0 & = & c^{2}dt^{2}\left(1-\left(\frac{d\tau}{dt}\right)^{2}\right)-dx^{2}_1-dx^{2}_2-dx^{2}_3\\
\iff\qquad 0 & = & v^{2}dt^{2}-dx^{2}_1-dx^{2}_2-dx^{2}_3\label{eq:gr-eq-47-1}
\end{eqnarray}
The novelty of the theorem is not that any freely moving particle with or without mass can be
 identified with motion along a null metric, although this too is important and follows directly from the  velocity
 equation $v^i=\frac{dx^i}{dt}$ as defined in the laboratory frame. Instead, the novelty rests on the fact that a
 necessary and sufficient condition for motion along a null metric to be transformed into a non-null metric in Minkowski
 space is that there should be an isotropic velocity $c$ valid for all frames of reference such that 
$\frac{d\tau^2}{dt^2}=1-\frac{v^2}{c^2}$ and $v\ne c$. In other words, the relationship between $dt$ and $d\tau$
 is defined by a Lorentz transformation. With this said, we now seek a spinor representation of the above theorem 
\\
\begin{lem}\label{lem:DE-2} If $ds^2=\eta_{ab}dx_adx_b$ then  $ds\xi=(c\gamma_0dt+\sum^3_{i=1} \gamma_idx^i)\xi \iff 0=(vdt+\sum^3_{i=1}\gamma^{\prime}_idx^i)\xi$ where
$$\gamma^{\prime}=\frac{c}{v}\left(\gamma_0+\frac{d\tau}{dt}\right),\  \gamma^{\prime}_i =
\frac{c}{v}\left(\gamma_0+\frac{d\tau}{dt}\right)\gamma_i, \qquad (\gamma^{\prime}_i)^2=1\ {\rm and} \{\gamma^{\prime}_i,\gamma^{\prime}_j\}=0 ,\ ds=cd\tau$$
\end{lem}
\noindent{Proof}: Since $ds^2=\eta_{ab}dx_adx_b$ the spinor square root exists such that \begin{eqnarray}ds\xi&=&(c\gamma_0dt+\sum \gamma_idx^i)\xi\qquad {\rm this\ is\ the\ Dirac\ square\  root}\\
\iff\ 0&=&[(c\gamma_0dt-ds)+\sum \gamma_idx^i]\xi \ {\rm subtract}\  ds\xi\ {\rm from\ both\ sides}\\
\iff\ 0&=&cdt\left(\gamma_0-\frac{d\tau}{dt}\right)\xi+\sum\gamma_idx^i\xi\ \ {\rm using}\ ds=cd\tau\\
\iff\ 0&=&\frac{c}{v}\left(\gamma_0+\frac{d\tau}{dt}\right)\left[cdt\left(\gamma_0-\frac{d\tau}{dt}\right)\xi+\sum\gamma_idx^i\xi\right]\ {\rm mult. by}\ \gamma^{\prime}\\
\iff\ 0&=&vdt\xi+\sum\gamma^{\prime}_idx^i\xi\qquad {\rm note}\ \left(\gamma_0+\frac{d\tau}{dt}\right)\left(\gamma_0-\frac{d\tau}{dt}\right)=\frac{v^2}{c^2}\label{eq:gr-eq-47-1a}
\end{eqnarray}
It is easy to check that for $1\le i \le 3$, $\gamma^{\prime 2}_i=-\gamma^2_i=1$, $\{\gamma^{\prime}_i,\gamma^{\prime}_j\}=0$ and $v^2dt^2\xi=\sum dx^2_i$
\\\newline
It should be clear in the above lemma that $\gamma_i$ and $\gamma^{\prime}_i$ are two different representations of the Dirac spin matrices which are dependent upon each other through the relationship $\gamma^{\prime}_i =
\frac{c}{v}\left(\gamma_0+\frac{d\tau}{dt}\right)\gamma_i$.  Once we have chosen a representation $\gamma_i$ to linearize the metric of Theorem \ref{thm:DE-1 } then the linearization of the null metric will depend on this choice of 
$\gamma_i$ or vice versa, as shown in the above lemma. We say more about the choice of $\gamma$ matrices in the next section. 
\begin{cor}\label{cor:DE-2}Given the relationship of Lemma (\ref{lem:DE-2})
there exists a wave equation given by
\begin{equation}\left(\frac{\partial \psi}{v\partial t}=-\sum^3_{i=1}(\gamma^\prime)^i\frac{\partial \psi}{\partial x^i}\right) 
\qquad {\rm where}\ \gamma^\prime_i\ \textrm{ is as in Lem(\ref{lem:DE-2})}. \label{eq:gr-eq-6-16}
\end{equation}    
\end{cor}
\noindent{Proof}: Take the dual of \ref{eq:gr-eq-47-1a}.
\\
\begin{cor}\label{lem:DE-3}
We can associate the spinor equation $vdt\xi=\sum^3_{i=1}\alpha_idx^i\xi$ with the metric 
$0=v^2dt^2-dx^2_1-dx^2_2-dx^2_3$ where $\alpha_i$ is a two dimensional representation of the spinor matrices.
Moreover, this can also be expressed in a four dimensional representation as $vdt=\sum\gamma^{\prime}_idx^i\xi$, where $\gamma_0$ is defined in Lem(\ref{lem:DE-2}).
\end{cor}
\noindent{\textbf {Proof}:} It follows from our initial assumptions that
\begin{eqnarray}0&=&v^2dt^2-dx^2_1-dx^2_2-dx^2_3\\
\iff\qquad v^2dt^2&=&dx^2_1+dx^2_2+dx^2_3
\end{eqnarray}
Since there are only three independent random variables, we can define a square root over a two dimensional spinor space with respect to a spinor basis $\alpha_i$ where $\alpha^2_i=1$  and $ \{\alpha_i,\alpha_j\}=0$ such that
\begin{equation}vdt\xi=\sum(\alpha_idx^i )\xi \end{equation}
The $\alpha_i$ can be imbedded in a four dimensional representation by defining
$$\gamma^{\prime}_i = \frac{c}{v}\left(\begin{array}{cc}
0 & \alpha_i+\alpha_i\frac{d\tau}{dt}\\ 
\alpha_i-\alpha_i\frac{d\tau}{dt} &0
\end{array}\right)$$
In which case \begin{equation}vdt\xi=\sum(\gamma^{\prime}_idx^i )\xi \end{equation}
\\
{\bf Remark}: We chose $\gamma^{\prime}_i$ in the above Lemma to be consistent with Lem(\ref{lem:DE-2}).
In fact, if we define $$\gamma_i = \left(\begin{array}{cc}
0 & \alpha_i\\ 
-\alpha_i & 0
\end{array}\right)\qquad \gamma_0 = \left(\begin{array}{cc}
I & 0\\ 
0 & -I
\end{array}\right)$$
then we see that $\gamma^{\prime}_i =
\frac{c}{v}\left(\gamma_0+\frac{d\tau}{dt}\right)\gamma_i$, as in the lemma.
In practice, the boundary or initial conditions will dictate the proper choice of  $\gamma_i$, and if no specific choice is required, the important thing will be to maintain consistancy throughout.
\begin{cor}\label{cor:DE-3}It follows from Cor(\ref{cor:DE-2}) that if $v$ is constant then
\begin{equation}0=\frac{1}{v^2}\frac{\partial^2 \psi}{\partial t^2}-\sum^3_{i=1}\frac{\partial^2 \psi}{\partial x^2_i}
\end{equation}
\end{cor}
\noindent{\textbf Proof}: It is sufficient to multiply equation (\ref{eq:gr-eq-6-16}) by 
$-\sum(\gamma^\prime)^i\frac{\partial \psi}{\partial x^i}$ or equivalently note that 
\begin{equation}0=\left(\gamma_0 \frac{\partial }{v\partial t}+\gamma_0\sum\gamma^{\prime}_i\frac{\partial }{\partial x^i}\right) \left(\gamma_0 \frac{\partial \psi}{v\partial t}+\gamma_0\sum\gamma^{\prime}_i\frac{\partial \psi}{\partial x^i}\right) =\frac{1}{v^2}\frac{\partial^2 \psi}{\partial t^2}-\sum\frac{\partial^2 \psi}{\partial x^2_i}\label{eq:gr-eq-48}
\end{equation}
\\

Both Cor(\ref{cor:DE-2}) and Cor(\ref{cor:DE-3}) express deep results. It means that the Dirac and 
the Klein-Gordan equations respectively can be reformulated in terms of classical wave equations defined locally 
over null geodesics  where 
the velocity of the particle is $v$. This $v$ is mass (and presumably charge) dependent. Indeed, from the perspective of special relativity, we can write 
$v^2=c^2\left(1-\left(\frac{m_0}{m}\right)^2\right)$. In other words, 
all wave equations for quantum particles are distinguished only by their velocity, which in turn will be dependent upon 
their mass and charge. The wave equation for em-radiation is a special case that occurs when $v=c$ or equivalently
 $m_0=0$. In the case of general relativity this mass-charge dependence is expressed in the Reissner-Nordstroem metric.

\subsection{Dirac equation and Clifford algebra properties}


Now consider the motion of a test particle of mass
$m$ along a timelike geodesic. Let $p^{a}=m(dx^{a}/d\tau)$, where
$\tau$ is the proper time (i.e. $ds=cd\tau$). Then 
\begin{equation}
ds^{2}=\eta_{ab}dx^{a}dx^{b}\qquad\mbox{is equivalent to}\qquad(mc)^{2}=\eta_{ab}p^{a}p^{b}.\label{eq:gr-eq-7-1}
\end{equation}
which in spinor notation means that 
\begin{equation}
ds\xi=\gamma_{a}dx^{a}\xi\mbox\ {\rm is\ equivalent\ to\ the\ Dirac\ Hamiltonian\ }\gamma^{a}p_{a}\xi(p)=mc\xi(p).\label{eq:gr-eq-8-1}
\end{equation}
If equation \ref{eq:gr-eq-6-1} is subjected to the constraints of
equation \ref{eq:gr-eq-8-1}, as it should be for motion along the
timelike geodesic, we find that $\psi=\psi^{i}(\int_{x_{0}}^{x(s)}p_{\mu}dx^{\mu})e_{i}$
is a solution of equation \ref{eq:gr-eq-6-1} in a general coordinate
system, provided the integration is taken along the curve $s=c\tau$
and $\xi(p)=\frac{d\psi^{i}(p)}{d\tau}e_{i}$. 

It is also worth noting that if all of $\psi^{i}$ are equal then
$\psi=\psi^{i}(x)\xi$ where $\xi$ is a spinor independent of $x$. In
this particular case the Dirac equation takes on the form 
\begin{eqnarray}
(\tilde{\partial}_{s}\psi^{i})\xi=\frac{\partial\psi^{i}}{\partial s}\xi.\label{eq:gr-eq-9-1}
\end{eqnarray}

Finally, we need to say something about the choice
of $\gamma$ matrices. Essentially we have defined them in terms of
the linearized metric $ds\xi=\gamma_{a}dx^{a}\xi$. In doing so, this
requires that $\{\gamma_{a},\gamma_{b}\}=2\eta_{ab}$. Note that there
are in infinite number of choices for these matrices. In Minkowski
space, it is conventional to choose 
\begin{equation}
\gamma^{0}=\left(\begin{array}{cc}
I & 0\\
0 & -I
\end{array}\right)\qquad\mbox{and}\qquad\gamma^{i}=\left(\begin{array}{cc}
0 & \alpha^{i}\\
-\alpha^{i} & 0
\end{array}\right),\ i\in\{1,2,3\} \ .
\end{equation}
where the $\alpha^{i}$'s are constant and correspond to the Pauli
spin matrices. However, if we switch to curvilinear coordinates then both
the $\gamma$ and $\alpha$ matrices may be no longer constant. For example if 
\[
ds^{2}=dx^{2}+dy^{2}
\]
then 
\[
\tilde{ds}=\alpha_{x}dx+\alpha_{y}dy
\]
where we choose 
\begin{equation}
\alpha_{x}=\left(\begin{array}{cc}
0 & 1\\
1 & 0
\end{array}\right)\qquad\mbox{and}\qquad\alpha_{y}=\left(\begin{array}{cc}
1 & 0\\
0 & -1
\end{array}\right).
\end{equation}
On switching to polar coordinates, $x=r\cos\theta$ and $y=r\sin\theta$
such that 
\[
\tilde{ds}=\alpha_{r}dr+\alpha_{\theta}rd\theta
\]
and 
\begin{equation}
\alpha_{r}=\left(\begin{array}{cc}
\sin\theta & \cos\theta\\
\cos\theta & -\sin\theta
\end{array}\right)\qquad\mbox{and}\qquad\alpha_{\theta}=\left(\begin{array}{cc}
\cos\theta & -\sin\theta\\
-\sin\theta & -\cos\theta
\end{array}\right).
\end{equation}
Clearly, $\alpha_{r}$ and $\alpha_{\theta}$ are now variables. In
the Heisenberg representation of quantum mechanics, operators are
variables that usually evolve in time. Specifically, when discussing
the zitterbewegung problem, the $\alpha$ oparators are variables
such that $i\dot{\alpha}=[\alpha,H_{o}]$, where $H_{o}=\alpha {\bf p}+\beta m$.
We will discuss this later on. 

\section{Hamilton-Jacobi functions and the Dirac equation}

To more fully appreciate wave-particle duality as defined by equations (\ref{eq:gr-eq-6-1}) and 
(\ref{eq:gr-eq-4-1}), we investigate their Clifford algebra properties. Recall that if $\bf{u}$ and 
$\bf{v}$ are two vectors with (symmetric) dot product $\bf{u}\cdot\bf{v}$
and (antisymmetric) cross product $\bf{u}\wedge\bf{v}$,
the Clifford product of $\bf{u}$ and $\bf{v}$ is defined by

\begin{equation}
\bf{u}\bf{v}=\bf{u}\cdot\bf{v} +\bf{u}\wedge\bf{v}.\label{eq:Clifford-product-rule}
\end{equation}

In the case of  the two operators $\tilde{ds}$ and $\tilde{\partial}_{s}$,
which are matrix representations of Clifford vectors their
Clifford product is 
\begin{eqnarray}
2\tilde{ds}\,\tilde{\partial}_{s}\psi & = & \{\tilde{ds},\tilde{\partial}_{s}\psi\}+[\tilde{ds},\tilde{\partial}_{s}\psi]\\
 & = & 2\frac{\partial\psi}{\partial s}ds+[\tilde{ds}\psi,\tilde{\partial}_{s}\psi]\label{eq:gr-eq-6a-1}
\end{eqnarray}
In contrast, if we multiply equations (\ref{eq:gr-eq-6-1}) and (\ref{eq:gr-eq-4-1})
together we obtain 
\begin{equation}
\tilde{ds}\,\tilde{\partial}_{s}\psi=\frac{\partial\psi}{\partial s}ds\ .\label{eq:gr-eq-6b-1}
\end{equation}
Equations \ref{eq:gr-eq-6a-1} and \ref{eq:gr-eq-6b-1} are compatible
if and only if $[\tilde{ds},\tilde{\partial}_{s}\psi]=0$. This is
important for many reasons:
\begin{enumerate}
\item In means that both $\tilde{ds}$ and $\tilde{\partial}_{s}\psi$ share
a complete set of common eigenvectors and therefore, from the perspective
of quantum mechanics, both are compatible and simultaneously measurable. 
\item Geometrically, it means that $\tilde{ds}$ and $\tilde{\partial}_{s}\psi$
are parallel at each point of the curve $s=s(t,x,y,z)$.  
\item $d\psi=\frac{\partial\psi}{\partial s}ds$ is an exact differential
and obeys the Hamilton-Jacobi equation. The precise meaning of this
point will be discussed below.
\item Historically, Schroedinger's work was motivated by the Hamilton-Jacobi
equation of classical mechanics \cite{oraf-wp},\cite{schr}. 
\end{enumerate}
For much of what follows, we will assume that $[\tilde{ds},\tilde{\partial}_{s}\psi]=0$.
This is not such a restrictive assumption as we shall see from Lemma
\ref{lem:HJ-2} below. Essentially, one can always decompose
$\psi$ into a parallel and perpendicular component given by $\psi(x)=\psi_{||}+\psi_{\bot}$.
Combined with the requirement of differentiability, this condition
means that $\psi_{\bot}$ is purely linear and consequently $[\tilde{ds},\tilde{\partial}_{s}\psi_{\bot}]=0$.
In other words, in the case of a real function, the requirement that
$\psi_{\bot}$ be an exact differential corresponds to tangential
motion with an intrinsic constant angular momentum along the curve.

If we return to equations (\ref{eq:gr-eq-6a-1}) and (\ref{eq:gr-eq-6b-1}),
there is another way of viewing the relationship between both expressions
in terms of Hamilton-Jacobi functions and the notion of exact differentials. In fact (\ref{eq:gr-eq-6b-1})
is an exact differential in that 
\begin{equation}
\tilde{ds}\,\tilde{\partial}_{s}\psi=\frac{\partial\psi}{\partial s}ds=\frac{\partial\psi}{\partial x^{a}}dx^{a}\ \label{eq:gr-eq-14-1}
\end{equation}
and as we shall see,  can be
associated with a coherent set of natural motions \cite{sg}. Indeed,
to remove any ambiguity, we begin with the following definitions:
\begin{defn}\label{defn:HJ}
A function $W=\int_{\sigma(\lambda)}\left({\mathbf{p}}{\frac{\mathbf{dx}}{d\lambda}}-H\frac{dt}{d\lambda}\right)d\lambda$
is called a Hamilton-Jacobi function if the integral is path independent
for all curves in $\{\sigma(\lambda)\in({M},g)\}$ and $\frac{\partial W}{\partial t}=-H(x_{1},x_{2},x_{3},t,\frac{\partial W}{\partial x_{1}},\frac{\partial W}{\partial x_{2}},\frac{\partial W}{\partial x_{3}})$
defined with respect to a local tetrad. Equivalently, we can say that $dW={\mathbf{pdx}}-Hdt$ is an exact differential.
\end{defn}

In relation to equations (\ref{eq:gr-eq-6a-1}) and (\ref{eq:gr-eq-6b-1}), we state (and prove) the following important theorem:

\begin{lem}
\label{lem:HJ-2}Let $\psi(W({\mathbf{x}},t))$ be a differentiable
function and $\{\sigma(\lambda)\}$ a family of curves on the manifold
with unit tangent vectors $\frac{\tilde{ds}}{ds}$ with respect to
a local tetrad then $\tilde{ds}.\tilde{\partial}_{s}\psi(W)$ is an
exact differential iff $\psi(W)$ is a Hamilton-Jacobi function such
that $p_{\| a}^{*}=\frac{d\psi}{ds}\frac{dx_{a}}{dt}=\frac{\partial\psi_{\|}(W)}{\partial x^{a}}$,
where $\psi(W)=\psi_{\|}(W)+\psi_{\bot}(W)$ and $\psi_{\bot}(W)=c_{o}t+c_{1}x_{1}+c_{2}x_{2}+c_{3}x_{3}$,
with $c_{o},c_{1},c_{2},c_{3}$ being constants.
\end{lem}

\noindent \textbf{Proof:} see appendix, lemma \ref{lem:HJ-2-app}.
\\

{\begin{cor}\label{cor:HJ-10} Let $\tilde{ds}.\tilde{\partial}_{s}\psi(W)$ be an
exact differential then
 $2\tilde{ds}.\tilde{\partial}_{s}\psi(W)=2\frac{\partial \psi}{\partial x^a}dx^a=\{\tilde{ds},\tilde{\partial}_{s}\psi(W)\}$

\end{cor} \textbf{Proof:} This follows both from equation(\ref{eq:Clifford-product-rule}) as noted in  lemma(\ref{lem:HJ-2}) and from the exact differentiability of $\tilde{ds}.\tilde{\partial}_{s}\psi(W)$ 
which means $[\tilde{ds},\tilde{\partial}_{s}\psi(W)]=0$
\newline

In effect, Lemma \ref{lem:HJ-2} establishes a relationship between
Hamilton-Jacobi functions and the commutator relationship, $[\tilde{ds},\tilde{\partial_{s}}\psi(W)]$.
We now use the same commutator relationship
to establish another important property relating the dual operator
$\tilde{\partial}_{s}$ and the metric operator $\tilde{ds}$ associated
with the increments along a curve.  Intuitively, we could think of $\tilde{\partial}\psi(W(s))$
as a wave associated with the vibration of a curve $\sigma(s)$ in
space-time, whose tangent is $\tilde{ds}$, with respect to a local
tetrad coordinate system. This leads to the following lemma: 
\begin{lem}
If $\psi(W)$ is a Hamilton-Jacobi function such that $[\tilde{\partial_{s}}W,\tilde{ds}]=0$
then there exists a simultaneous eigenfunction $\xi$ such that 
\begin{equation}
(\tilde{\partial_{s}}\psi)\xi(p)=\partial_{s}\psi\xi(p),\label{eq:gr-eq-33-1}
\end{equation}
where $\partial\psi_{s}=\frac{\partial\psi}{\partial s}$, which in
the case of geodesic motion reduces to 
\begin{equation}
\tilde{\partial_{s}}\Psi=\frac{d\Psi}{ds},\qquad\textrm{where}\qquad\Psi=\psi\xi.\label{eq:gr-eq-34-1}
\end{equation}
\textbf{Remark:} $(\tilde{\partial_{s}}\psi)\xi=\tilde{\partial_{s}}\Psi$
in general, since $x$ is independent of $p$ in phase space. However,
$\frac{d\Psi}{ds}\neq\psi^{\prime}(p)\xi$ unless motion is along
a geodesic. 
\end{lem}

\textbf{Proof:} First note that $[\tilde{\partial_{s}}W,\tilde{ds}]=0$
implies $[\tilde{\partial_{s}}\psi,\tilde{ds}]=[\psi^{\prime}(W)\tilde{\partial_{s}}W,\tilde{ds}]=0$.
Therefore, there exists simultaneous eigenvectors $\xi=\xi(p)$ such
that $\tilde{ds}\xi=ds\xi$ and $(\tilde{\partial_{s}}\psi)\xi=\gamma^{a}p_{a}^{*}=\gamma^{a}p_{a}\psi^{\prime}\xi(p)=mc\psi^{\prime}(p)\xi(p)=(\partial_{s}\psi)\xi(p)$.
Also, $\xi(p)$ is constant along a geodesic and therefore 
\[ 
\tilde{\partial_{s}}\Psi=\frac{d\Psi}{ds},\qquad\textrm{where}\qquad\Psi=\psi\xi.
\]
The result follows.

\noindent \textbf{Remark:} We refer to Equation (\ref{eq:gr-eq-33-1})
as a generalized Dirac equation associated with a curve, and (\ref{eq:gr-eq-34-1})
as a generalized Dirac equation associated with geodesics. Both of these are subsumed by 
equation(\ref{eq:gr-eq-6-1}) which we have also referred to as a generalized Dirac equation. It reduces
to the usual form of the Dirac equation if we let $\psi=Ae^{\Lambda W}$,
where $A$ is an arbitrary constant and $\Lambda=\frac{i}{\hbar}$.\\
\noindent 

\subsection{Exact differentials and metrics}
In this section, we investigate the relationship between the Hamilton-Jacobi function $W$ considered as an 
exact differential and its relationship to a family of metrics, $ds$. Specifically, since $dW$ is exact then 
$p_a=\frac{\partial W}{\partial x^a}$  depends explicitly only on the coordinates and not on 
the parametrizaion per se. Therefore, to be consistent we are required to choose a $\lambda$ in the 
definition \ref{defn:HJ} above, so that $p_a$ is invariant with respect to
some agreed (universal) standard parameter, which emerges naturally from the geometrical and dynamical 
requirements of the system. For example, following the usual rule of dynamics, we require that in the instantaneous 
rest frame 
$$p_a =m(s)c\frac{dx^a}{ds}$$
where $m(s)$ is the instantaneous mass on a specific curve, parametrized with respect to its proper time $ds$. 
Note that $m(s)$ is not necessarily a constant except along a geodesic. Moreover, if we were 
to change the parameter from $s$ to some other  $s^{\prime}$ then for $p_a$ to be
invariant at each point with respect to its coordinate system would require that
\begin{equation} p_a=m(s)c\frac{dx^a}{ds}=m(s^{\prime})c\frac{dx^a}{ds^{\prime}}\ {\rm with}\ 
\frac{ds^{\prime}}{m(s^{\prime})}\equiv \frac{ds}{m(s)}\end{equation}
This suggests that in a gravitational field we should define $d\lambda=\frac{ds}{m(s)}$ as a 
universal parameter with respect to some standard particle. Note in the laboratory frame, 
we would have $\frac{ds}{m(s)}=\frac{dt}{m(t)}$ \cite{poh-cm}. Other refences to this particular parametrization can also be found in \cite{jc-ca} and \cite{lh-re}.

With that said, we now address the important question of which comes first, the Hamilton-Jacobi function or the metric associated with the dynamics of the particle.  In reality it does not matter provided one or the other is an exact differential, although  
there are two points to keep in mind: 
\begin{enumerate}
\item Given an arbitrary metric  $ds^2=dx_adx^a$ does not necessarily mean that we can construct a 
Hamilton-Jacobi function, $W$, from it. More precisely, given a metric $ds$ there does not necessarily exist an integrating 
factor $\rho=m(x^a)c$ such that 
$dW=\rho ds$ is an exact differential. For example if 
$$ds=-ydx+xdy+kdz$$
and we require $dW=\rho(x,y,z)ds$ then on solving we find that $\rho=0$ . On the other hand if we let
$$ds=\frac{yz}{\rho}dx+\frac{xz}{\rho}dy+\frac{xy}{\rho}dz$$
then for all $\rho(x,y,z)\ne \rho(xyz)$, $ds$ is not exact although $dW=\rho ds$ is always exact. 

\item  Given an exact differential $dW$ can we associate a metric with it? The answer is yes: 
\begin{thm}Let $dW$ be an exact differential such that $dW=\frac{\partial W}{\partial x^a}dx^a$ then a family 
of curves with metric  $ds^2=dx_a dx^a$ exists such that $dW=\rho ds$, where $\rho$ is an integrating factor such that
$\rho^2=\frac{\partial W}{\partial x^a}\frac{\partial W}{\partial x_a}$   .\newline

\noindent{\bf Proof:} Note that if there exists a metric $ds$ such that $dW=\rho ds$ then one can also define 
$ds_1=\Lambda ds$ and $\rho_1=\frac{\rho}{\Lambda}$ such that $dW=\rho_1ds_1$, indeed, one can choose 
$\rho_1=1$. This presents us with the question of choosing a standard (universal) parameter for the system. For what follows, we choose $ds$ such that  
$dx^a/ds$ is the unit tangent vector along the a family of regular curves (meaning $ds\ne 0$) on some interval of $s$.  Since $dW$ is exact the value of $W(x_a)$ is independent of the path, and therefore the choice of $\rho$ 
will depend on the parametrized coordinate system (see below) . In fact,
\begin{eqnarray} dW &=&\rho ds\\
&=&\rho\left(\frac{dx_a}{ds}dx^a\right)\\
&=&\left(\rho\frac{dx_a}{ds}\right)dx^a\\
&=&\frac{\partial W}{\partial x^a}dx^a\qquad {\rm since}\ dW\ {\rm is\ exact}
\end{eqnarray}
Therefore,
$$\frac{\partial W}{\partial x^a}=\rho \frac{dx^a}{ds}$$ 
But by definition of $ds^2$, we have that $\frac{dx_a}{ds}\frac{dx^a}{ds}=1$ and consequently
$\rho^2=\frac{\partial W}{\partial x_a} \frac{\partial W}{\partial x^a}$ and invariant under Lorentz transformations.
\end{thm}
\end{enumerate}
\begin{description}
\item[Remark 1] In the case of a Hamilton-Jacobi function $W$, we note that the integrating factor $\rho$ has the units of momentum. Indeed, if we let $d\lambda=ds/m(x^a)$  be the universal parameter as suggested above then 
$\rho=m(x^a)c$ along the curve.  

\item[Remark 2] If $\rho= \frac{dW(s)}{ds}$ then by the chain rule both $dW$ and $ds$ are exact. 
To see this it is sufficient to note that $dW=\frac{dW}{ds}ds$ and $ds=\frac{ds}{dW}dW$.

\item[Remark 3] In the case of thermodynamics  $W$ would be entropy, $s$ the total heat energy and $\rho=1/T$ where $T$ is temperature.
\end{description}

\subsection{Some examples}\label{SE}
The above is subtle. It does not claim that every metric can be associated with an exact differential but the 
opposite, namely that every exact differental can be associated with a metric and consquently a specific family of curves.  We give three examples:
\begin{itemize} 
\item Consider a metric given locally by $ds^2=dx_adx^a$ such that
$$s=\cos(\theta_o)t-i\cos(\theta_j)x^j,\ j\in\{1,2,3\}$$ and $\cos{\theta_a}$ a constant. $s$ is clearly exact. Consequenly, 
we may take $W=m_ocs$ and $\rho=m_oc$, which follows by applying the theorem directly to obtain 
$$\rho^2=\frac{\partial W}{\partial x_a}\frac{\partial W}{\partial x^a}=(m_oc)^2\cos(\theta^a)\cos(\theta^a)=
(m_oc)^2$$ 
Note $\frac{\partial s}{\partial x^a}=\frac{dx^a}{ds}=\cos(\theta^a)$ by definition of directed cosines on a local Minkowski space. It is easy to check that $\nabla (s).\vec{dx}=ds$.
In this case $W$ is a Hamiltonian function along a geodesic.
\item As a second example consider $W=xy$ then we have
\begin{eqnarray}dW&=&\frac{\partial W}{\partial x}dx+\frac{\partial W}{\partial y}dy\\
&=&ydx+xdy
\end{eqnarray}
From the above theorem we obtain,
$$\rho^2=\frac{\partial W}{\partial x_a}\frac{\partial W}{\partial x^a}=y^2+x^2$$
\begin{equation} \frac{\partial W}{\partial x}=y=\rho\frac{dx}{ds}\end{equation}
and
\begin{equation} \frac{\partial W}{\partial y}=x=\rho\frac{dy}{ds}\end{equation}
This means that
\begin{equation}\frac{dy}{dx}=\frac{x}{y}\qquad {\rm and}\qquad  y^2-x^2=\kappa^2\end{equation}
which defines a family of hyperbolae.
Therefore, the standard metric associated with $W$ is given by
$$ds=\frac{y}{\sqrt{x^2+y^2}}dx+\frac{x}{\sqrt{x^2+y^2}}dy$$
Note that $ds$ is not an exact differential but $\sqrt{(x^2+y^2)}ds$ is exact.

\item As a third example, we define $W=x^2+y$ then $dW=2xdx+dy$. From the above theorem we obtain,
\begin{equation} \frac{\partial W}{\partial x}=2x=\rho\frac{dx}{ds}\end{equation}
and
\begin{equation} \frac{\partial W}{\partial y}=1=\rho\frac{dy}{ds}\end{equation}
This means that
\begin{equation}\frac{dy}{dx}=\frac{1}{2x}\iff  y=\ln\sqrt{x} + \kappa,\ {\rm and}\ \lambda^2= 
\left(\frac{\partial W}{\partial x}\right)^2+  \left(\frac{\partial W}{\partial x}\right)^2=4x^2+1 \end{equation}
which defines an exponential family of curves given by $x=A\exp(2y)$. 
\end{itemize}

\section{Different representations of the wave function and the zitterbewegung}

In classical mechanics, we have essentially three orders of equations. The first describes the particle trajectory and can 
be represented by a parmetrized curve $x^a=x^a(\tau)$, where $\tau$ is a parameter. 
The second level corresponds to
kinematics with $p^a=m\dot{x}^a$, where $\dot{x}^a=\frac{dx^a}{d\tau}$ and the third is given by Newton's equation
 of motion $\dot{p}^a=m\ddot{x}^a$. Essentially, in a deterministic system if one is known the other two can be
 calculated.  In Hamiltonian
 mechanics, there is an analogous structure which is summarized in 
the table below by associating the curve and the action, $s=(x^a)\longleftrightarrow W$ as follows:

\noindent\fbox{\begin{minipage}[t]{1\columnwidth - 2\fboxsep - 2\fboxrule}%
\[ s(\tau)=(x^a(\tau)) \longleftrightarrow W=W(x^a)
\]

\[
	ds^{2}=\eta_{ab}dx^{a}dx^{b}\:\longleftrightarrow\:dW=\frac{\partial W}{\partial x^a }dx^a
\]

\[
p^a=m\dot{x}^a \longleftrightarrow p^a=-\eta^{ab}\frac{\partial W}{\partial x_b},\  p^0(t)=H(t)=H(\tau)\dot{t},\ \frac{\partial\psi(W)}{\partial s}=\gamma^{a}\frac{\partial\psi(W)}{\partial x^{a}}
\]

\[
\dot{p}^a =m\ddot{x}^a\longleftrightarrow \dot{p}^a=\eta^{ab}\frac{\partial H(\tau)}{\partial x_b}, \ 
\dot{x}^a=\eta^{ab}\frac{\partial H(\tau)}{\partial p_a}
\]

%
\end{minipage}}
\\
\\
What is important to note from the above table is that the generalized Dirac ``wave equation" (\ref{eq:gr-eq-6-1})
is essentially associated with the kinematics of a particle and not directly with its equations of motion.
Indeed, when motion is along a geodesic, the essential information is captured by the kinematics. 
This explains one reason why on solving the Dirac (Schroedinger) equation the solutions are associated with 
fixed energy-momentum levels, which by definition pertain to geodesics. Furthermore, we put the expression
``wave-equation'' in quotes to emphasize that $\psi(W)$ is not
necessarily a wave-function of quantum mechanics. For the moment,
$\psi$ can be any $C_{1}$ function defined on the manifold. $\psi$
should only be interpreted as a quantum wave state when further restrictions
are imposed on the function space, such as requiring that it be an element of a Hilbert space
$L^{2}$ and that Planck's equation, $E=h\nu$ relates the frequency of an oscillator and/or particle to its energy. We now present some examples to help clarify the above theory.

\subsection{Wave-particle duality and the zitterbewegung} 
To analyze the motion of a quantum particle, we need to keep in mind that unlike the classical case the quantum particle is subjected to quantization conditions or equivalently the Heisenberg uncertainty relations, which can be derived when the wave function is not a point mass (as in the classical case) but smooth and continuous. 
Indeed, regular quantum mechanics presupposes the wave function to be an element of a Hilbert space that obey continuity conditions at the boundaries which corresponds to standing wave solutions for atomic oscillators. The energy of these standing waves are related by the 
Planck-Einstein formula given by $E=h\nu$, 
where $h$ is Planck's constant and $\nu$ is a frequency of an atomic oscillator or equivalently using the 
de Broglie relation $p=h/\lambda$, where $\lambda$ corresponds to the particle wavelength and $h$ 
is a constant of proportionality that in effect behaves as a scaling factor for measuring energy and momentum. 
$h$ takes on different values according to the system of units that is been used. Nowadays, it is not uncommon to 
take $h=1$ in natural units. However, it is also important to note that $h$ can never be 0.

The second point is that in the case of a Hilbert space, or more precisely $L^2$ functions, the inner product
$\left<\psi|\psi\right>$ not only has a statistical interpretation but is gauge invariant with respect to phase factors. 
This means that if 
$\left|\psi \right>= e^{iW}\left|\xi \right>$ then   
$\left<\psi|\psi\right>=\left<\xi|\xi\right>$ is invariant for all $W$. Note that if the $i$ were dropped then $\left<\psi|\psi\right>$ would not necessarily be invariant. 

This brings us to a third point. In conventional quantum mechanics the momentum operator is usually defined by $p^{\mu}\psi(W)=-i\hbar\frac{\partial \psi}{\partial x^{\mu}}$ which essentially means a rescaling of the Hamilton-Jacobi relation $p^a=\frac{\partial W}{\partial x^a}$, with the $i$ being associated with the phase factor noted in the previous paragraph. This effectively changes nothing from the prospective of 
the generalized Dirac equation in that 
$$\frac{\partial\psi}{\partial s}=\gamma^{a}\frac{\partial\psi}{\partial x^{a}} \iff i\hbar\frac{\partial\psi}{\partial s}=i\hbar\gamma^{a}\frac{\partial\psi}{\partial x^{a}}\ .$$
However, to fully exploit the gauge invariance and the de Broglie relations mentioned above, when we refer to the original Dirac equation along a geodesic,
convention has it that $i\hbar\frac{\partial\psi}{\partial s}=mc\psi$ where $\psi=\exp(-(i/\hbar)mcs)\xi$. 
In other words, the requirement of gauge invariance and that the eigenvalues be real  suggests redefining the 
Hamilton-Jacobi operators by multiplying them by an $i\hbar$ term. The important thing is to be consistent throughout.   
Therefore, we define ${\bf p_a}=i\hbar\partial_a$ with the understanding that $p_a=i\hbar\partial_aW$ from the Hamilton-Jacobi function. 
Note that this implies ${\bf p^a}=\eta^{ab}{\bf p_b}$ which means that 
${\bf p}^0={\bf p}_0$ but ${\bf p}^1=-i\hbar \partial_1$, ${\bf p}^2=-i\hbar \partial_2$ and 
${\bf p}^3=-i\hbar\partial_3$.
Indeed, if $W=mc^2 \tau$ (recall $s=c\tau$) then we can write  $$H_0=\frac{\partial W}{\partial \tau}=mc^2\qquad {\textrm and}\qquad 
H^*_0\equiv \frac{\partial{e^W}}{\partial \tau}=mc^2e^W$$
However, if we replace $W$ with $(-i/\hbar) W$ and write the Hamilton-Jacobi operator as $i\hbar \frac{\partial }{\partial \tau}$ 
then we find that once again, 
 $$H_0=i\hbar\frac{\partial (-i/\hbar)W}{\partial \tau}=\frac{\partial W}{\partial \tau}=mc^2\ {\textrm and} \
H^*_0\equiv i\hbar\frac{\partial e^{(-i/\hbar)W}}{\partial \tau}=mc^2e^{(-i/\hbar)W}$$

To conclude, the generalized Dirac equation is a Hamilton-Jacobi equation both for classical (relativistic) mechanics and quantum mechanics. In the classical
case the line increment from which the Hamilton-Jacobi function is
derived gives rise to a dual non-quantum ``wave-function,'' which
is a point mass. In contrast, in
the case of quantum mechanics the same line increment is dual to a
family of $L^{2}$ functions in such way that the initial boundary
conditions coming from the physics are statistical, non-deterministic and
incorporates quantization by seeking standing wave solutions. In other words, the mechanics
of a strictly classical particle can be determined (in principle)
from the initial conditions applied directly to the properties of
the line increments, with the non-quantum ``wave-equation'' representing
a point-mass and not contributing any additional information. In the
case of a quantum particle, because of the Heisenberg uncertainty
principle, the opposite appears to be true. It is precisely the ``wave-equation''
that encapsulates the kinematics of the particle, although the solution
to the ``wave-equation'' is dependent upon the line increments associated
with the classical particle. This also gives a new insight into the
Principle of Complementarity, in that the general solution to the Dirac wave equation is composed of a point 
mass solution ($\delta$ function)  plus a wave solution (an $L^2$ function). This in turn, enables us to have a more 
complete understanding of the zitterbewungung problem}

\subsection{The zitterbewegung problem:}Consider a particle of rest mass $m_{o}$
moving with uniform velocity $u$ with respect to proper time
along the x-axis in Minkowski space. This also means that the motion
of the particle with respect to two different frames in uniform motion
relative to each other are related by Lorentz transformations. The
Hamilton-Jacobi function of the particle is given by $W=m_{o}ux-m_{o}c^2\dot{t}t=-m_{o}c^2\tau$
such that $H=m_{o}c^{2}\dot{t}$. In the rest
frame, when $t=\tau$, we have $\dot{t}=1$ and consequently $H_{0}=m_{0}c^{2}$.
Indeed as a classical (relativistic) particle with $x=0$
when $\tau=0$, then $x=u\tau$, where $u=\frac{dx}{d\tau}$. In terms
of the coordinate system $(x,t)$ of the laboratory frame this can
be written as $x=vt$, where $v=(u\frac{d\tau}{dt})$ is
constant and obeys the wave equation
$$\frac{1}{v^2}\frac{\partial^2 \psi}{\partial t^2}-\frac{\partial^2 \psi}{\partial x^2}=0$$ 
associated with a null metric $0=v^2dt^2-dx^2$ (cf. Cor (5)).
The general soluton of this equation is given by $\psi(t,x)=Af(x-vt)+Bg(x+vt)$. 
However, from the initial conditions, and the fact that we are describing a point ``quantum'' particle,
we obtain a specific solution of the form  $\psi(x,t)=\delta_k(x-vt)$, $k$ a constant associated with the center of mass.

This information can also be derived by noting that the Hamilton-Jacobi function for a null metric is given 
by  $W_o=m_ock$ an unknown constant where $W_o$ refers to the Hamilton-Jacobi function associated with the null metric. Therefore, for a point particle the general solution is the form $\psi(W_o)=\delta_k$, where $k$ represents the point mass of the delta function. 
But $x-vt=k$ for a particle moving with uniform velocity $v$ which means $\psi(W_o)=\delta_k(x-vt)$. 
It might be worth noting that $W$ and $W_o$ are related by the following identities 
$$\frac{\partial W_o}{\partial t}=\frac{v^2}{c^2}\frac{\partial W}{\partial t}\qquad 
\frac{\partial W_o}{\partial x}=\frac{\partial W}{\partial x}\ $$ or in other words, $W_o$ is 
derived from $W$ by rescaling the local time variable.

In physical terms, this means that the particle's center of mass moves in a noise-free
environment. The wave functional per se adds no new information. For elementary particles like electrons, 
such initial conditions are unknown in principle because of 
the uncertainty relations. Indeed, precisely because the initial position is unknown, at best we can 
write $\psi(x,t)=\delta_k(x-vt)$, where $k$ is some unknown constant, which also means that 
while motion might be deterministic in principle, in practice (and in principle) it is unknowable. 
We can only hope to glean more information by way of probabilistic methods which are not a cloak for
 ignorance but rather capture the ontological reality of the zitterbewegung problem.  

It also stongly indicates that an 
elementary particle is not a particle in a classical sense and appears to have some structure implicitly suggested 
by combining  equations (\ref{eq:gr-eq-47-1}) and (\ref{eq:gr-eq-48}) with the zitterbewegung. They suggest that a free falling particle
obeys the same equation as a transversal electromagnetic wave, but
with a different velocity such that $v=v(m)$ with $v(0)=c$. In terms of quantum mechanics,
it is as if wave-particle duality corresponds to an entrapped electromagnetic
wave with energy given by $E=mc^{2}=nh\nu$, with each elementary
particle being characterized by its own distinct $\nu$. As we shall now see, this latter characteristic offers one explanation of the zitterbewegung effect

In contrast to the classical problem, the Heisenberg uncertainty relations, which are given by $\Delta x\Delta p\ge\hbar/2$
and $\Delta E\Delta t\ge\hbar/2$, change things radically and suggest that there is an underlying interference due to background radiation and/or the presence of electromagnetic noise interacting with the particle structure.
Indeed, the whole study of decoherence related to entanglement is another indication that the background noise is a 
determinig factor in particle motion. In that regard, the act of measurement is 
just another form of interference which is also subjected to the uncertainty relations. 
And these relations, combined with De Broglie's formula, lead to quantum mechanics. 

To better understand this, let us restrict energy exchanges
between particles to be a multiple of $nh$ and write $W+\Delta W=W+nh$
(\cite{sg}, p.461). This leads to the simple formula: 
\begin{equation}
(W+\Delta W-W)=\Delta W=nh
\end{equation}
from which it follows that in the rest frame of a free particle, we have
\begin{equation}
\frac{dW}{d\tau}=H_{0}\approx\frac{\Delta W}{\Delta\tau}=\frac{nh}{\Delta\tau}=nh\nu_{o},\qquad{\rm where}\qquad\nu_{o}=1/\Delta\tau\ .\label{eq:De-Broglie-from-W}
\end{equation}
This means that every elementary particle with rest mass $m_0$ can
be characterized by a standard frequency (wavelength) given by the
equation $m_0c^{2}=h\nu_{o}$. This is a consequence of  De Broglie's formula.

In this case, because of the uncertainty principle, the particle characteristics
are embedded within the wave function and not vice-versa. This can
be especially seen in the zitterbewegung effect.  As previously noted,  the position of a particle constrained
to move on the line is unknown because of the uncertainty relations.
The best we can do is describe the position by means of a uniform
density function $f(x,t)=1/\xi$ for $x\in[0,\xi]$ and introduce
a wave-function on a Hilbert space whose inner product gives the probability
distribution. We associate the Hamilton-Jacobi equation not with the probability $f$ but with the 
wave-fuction $\psi(W)$ and also encapsulate Planck's constant into $\psi$. 
In other words, although
$f=\left<\psi|\psi\right>$, we should not overlook the fact that $\psi$ (and not $f$) is the quantum wave
function. Also, based on the wave-particle duality properties previously developed, this suggests writing the 
wave function for a free particle (constant $p^{a}$) in spinor notation as a self adjoint eigenfunction such that 
$\psi(W)=\psi(\int p^{a}dx_{a})$. 
Moreover, in order for our notation to be consistent with the traditional Heisenberg approach 
of solving the equation $i\dot{\alpha}=[\alpha,H]$ to explain the zitterbewegung problem, 
we note by corollary(\ref{cor:HJ-10}) that we can use inner product notation, 
analogous to Heisenberg's outer product notation, 
to write $2\int(p_adx^a)=\{\tilde{\partial_{s}}\psi(W),\tilde{ds}\}$. It follows that:
\begin{eqnarray}
\qquad\psi & = &\psi(\int \{\tilde{\partial_{s}}\psi(W),\tilde{ds}\})\\
 &=& \psi^{i}(2\int^{x}p^{a}dx_{a})e_{i}\\
 & = & \frac{\exp(2i\hbar^{-1}H_{o}s)}{\sqrt{\xi}}\psi_{o}\\
 & = & \frac{\exp(2i\hbar^{-1}m_0c^{2}\tau)}{\sqrt{\xi}}\psi_{o}\\
 & = & \frac{\exp(2i\hbar^{-1}m_0c\lambda)}{\sqrt{\xi}}\psi_{o}
\end{eqnarray}

This zitterbewegung structure can be associated with a periodic (simple
harmonic) isotropic vibration of a particle with rest energy $m_0c^{2}$
and with De Broglie wavelength $\lambda$. It should also be noted
that the above formula contains no mention of electric charge. 

{\bf Remark:} In many textbooks, the zitterbewungung problem is analyzed 
from the perspective of the 
Heisenberg equation of motion $i\dot{\alpha}=[\alpha,H_0]$, for the ``velocity operator'' $\alpha$ and the 
free particle Hamiltonian, $H_0=\alpha{\bf p}+\beta m$. This equation of motion has the formal solution \cite{pm}:
$$x(t)=k+vt-ke^{-2iH_ot/\hbar}$$
This contains the linear part that corresponds to the particle motion and expressed by $x(t)=k+vt$ and the wave 
part (the zitterbewegung effect) expressed by the $e^{-2iH_ot/\hbar}$. However, the presentation given above 
using a Schroedinger/Dirac approach is more intuitive and helps us better understand the wave-particle properties implicit in the Heiesenberg approach. Moreover, as already pointed out in section 3, by taking $\psi_{\bot}=k+vt$, permits us in the light of Lemma \ref{lem:HJ-2} to also interpret this as an extra term corresponding to a free particle in motion but with a constant spin angular momentum.

\subsection{Summary:} The previous sections can be summarized
by the formulas listed below. Wave-particle duality corresponds to
the relationship between the metric and its assoociated spinor $\xi$ and a quantum mechanical wavefunction equation
which is the dual of a metric with wave function $\psi$ which we have applied to give a more intuitive understanding of the zitterbewegung frequency .

\noindent\fbox{\begin{minipage}[t]{1\columnwidth - 2\fboxsep - 2\fboxrule}%
\[
ds^{2}=g_{\mu\nu}dx^{\mu}dx^{\nu}=\eta_{ab}dx^{a}dx^{b}\:\longleftrightarrow\:ds\xi=\gamma_{a}dx^{a}\xi
\]

\[
\frac{\partial^{2}\psi}{\partial s^{2}}=\eta_{ab}\frac{\partial^{2}\psi}{\partial x^{a}\partial x^{b}}\:\longleftrightarrow\:\frac{\partial\psi}{\partial s}=\gamma^{a}\frac{\partial\psi}{\partial x^{a}}
\]

\[
\frac{\partial\psi}{\partial s}=\gamma^{a}\frac{\partial\psi}{\partial x^{a}}
\]

\[
d\psi=\frac{\partial\psi}{\partial x^{a}}dx^{a}
\]
%
\end{minipage}}
\noindent\fbox{\begin{minipage}[t]{1\columnwidth - 2\fboxsep - 2\fboxrule}%
\[
ds^{2}=\eta_{ab}dx^{a}dx^{b}\:\longleftrightarrow\:0=v^2dt^2-dx^2_1-dx^2_2-dx^2_3 
\]

\[
\frac{\partial^{2}\psi}{\partial s^{2}}=\eta_{ab}\frac{\partial^{2}\psi}{\partial x^{a}\partial x^{b}}\:\longleftrightarrow\:0=\frac{1}{v^2}\frac{\partial^2\psi}{\partial t^2}-\frac{\partial^2\psi}{\partial x^2_1}-\frac{\partial^2\psi}{\partial x^2_2}-\frac{\partial^2\psi}{\partial x^2_3}
\]

\[
ds\xi=\gamma_{a}dx^{a}\xi \longleftrightarrow\:0=v\gamma_{0}dx^0+\gamma_{1}dx^1+\gamma_{2}dx^2+\gamma_{3}dx^3
\]

\[
\frac{\partial\psi}{\partial s}=\gamma^{a}\frac{\partial\psi}{\partial x^{a}}
\longleftrightarrow\:0=\frac{1}{v}\gamma^0\frac{\partial \psi}{\partial t}+
\gamma^1\frac{\partial \psi}{\partial x_1}+\gamma^2\frac{\partial \psi}{\partial x_2}+\gamma^3\frac{\partial \psi}{\partial x_3}
\]
%
\end{minipage}}
\\

\appendix
\section{Hamilton-Jacobi functions and exact differentials} 
\setcounter{section}{1}

We now give a proof  of lemma \ref{lem:HJ-2} in section 
\ref{sec:Metrics-Dirac-eq}which gives a necessary and sufficient
condition for the existence of such Hamilton-Jacobi functions associated with a generalized Dirac equation:\\
\begin{lem}
\label{lem:HJ-2-app}Let $\psi(W({\mathbf{x}},t))$ be a differentiable
function and $\{\sigma(\lambda)\}$ a family of curves on the manifold
with unit tangent vectors $\frac{\tilde{ds}}{ds}$ with respect to
a local tetrad then $\tilde{ds}.\tilde{\partial}_{s}\psi(W)$ is an
exact differential iff $\psi(W)$ is a Hamilton-Jacobi function such
that $p_{\|a}^{*}=\frac{d\psi}{ds}\frac{dx_{a}}{dt}=\frac{\partial\psi_{\|}(W)}{\partial x^{a}}$,
where $\psi(W)=\psi_{\|}(W)+\psi_{\bot}(W)$ and $\psi_{\bot}(W)=c_{o}t+c_{1}x_{1}+c_{2}x_{2}+c_{3}x_{3}$, with
$c_{o},c_{1},c_{2},c_{3}$ being constants.. 
\end{lem}

\textbf{Proof:} From equations (\ref{eq:gr-eq-6a-1}), if $\frac{\tilde{ds}}{ds}\tilde{\partial}_{s}\psi(W)$
is an exact differential then $[\frac{\tilde{ds}}{ds},\frac{\tilde{\partial}\psi(W)}{\partial s}]$
is also an exact differential. This means that $\psi(W)=\psi_{\|}(W)+\psi_{\bot}(W)$
where $[\frac{\tilde{ds}}{ds},\frac{\tilde{\partial}\psi_{\|}(W)}{\partial s}]=0$
and $\psi_{\bot}(W)=c_{o}t+c_{1}x_{1}+c_{2}x_{2}+c_{3}x_{3}$, with
$c_{o},c_{1},c_{2},c_{3}$ being constants.\\
 Also, $[\frac{\tilde{ds}}{ds},\frac{\tilde{\partial}\psi_{\|}(W)}{\partial s}]=0$
means $\frac{{dx^{a}}}{d\lambda}$ is parallel (for any parameter
$\lambda$ including the curve length $s$) to $\frac{\vec{\partial}{\psi}_{\|}(W)}{\partial s}$,
and $\exists m(\lambda)$ such that $p_{\|a}^{*}\equiv m(\lambda)c\frac{dx_{a}}{d\lambda}=\frac{\partial\psi_{\|}(W)}{\partial x^{a}}$ (where $c$, the velocity of light,  is a scaling factor).\\

\noindent Also given $\frac{\tilde{ds}}{ds}\tilde{\partial}_{s}\psi_{\|}(W)$
is an exact differential and denoting $x_{0}=t$, gives 
\begin{eqnarray*}
\frac{\tilde{ds}}{ds}\tilde{\partial}_{s}\psi_{\|}(W) & = & \frac{\partial\psi_{\|}(W)}{\partial x^{1}}\frac{dx^{1}}{ds}+\frac{\partial\psi_{\|}(W)}{\partial x^{2}}\frac{dx^{2}}{ds}+\frac{\partial\psi_{\|}(W)}{\partial x^{3}}\frac{dx^{3}}{ds}+\frac{\partial\psi_{\|}(W)}{\partial t}\frac{dt}{ds}\\
 & = & \frac{d\psi_{\|}(W)}{ds}.
\end{eqnarray*}
On substituting $m(s)c\frac{dx_{a}}{ds}=\frac{\partial\psi_{\|}(W)}{\partial x^{a}}$
and noting that $\frac{dx^{a}}{ds}\frac{dx_{a}}{ds}=1$ gives\\
$m(s)c=\frac{d\psi_{\|}(W)}{ds}$. It follows that 
\begin{equation}
\left(\frac{\partial\psi_{\|}(W)}{\partial t}\right)^{2}=(p_{\|}^{*1})^{2}+(p_{\|}^{*2})^{2}+(p_{\|}^{*3})^{2}+\left(\frac{d\psi_{\|}(W)}{ds}\right)^{2}=\left(H_{o}^{*}\left(x^{a},p_{\|}^{*1},p_{\|}^{*2},p_{\|}^{*3}\right)\right)^{2}\label{eq:gr-eq-20-1}
\end{equation}
Therefore $\psi_{\|}(W)$ is a Hamilton-Jacobi function, as also is
$\psi(W)=\psi_{\|}(W)+\psi_{\bot}(W)$ with $p_{a}^{*}=p_{\|a}^{*}+c_{a}$.
\\

\noindent Conversely, given $p_{a}^{*}=\frac{d\psi}{ds}\frac{dx_{a}}{dt}+c_{a}=\frac{\partial\psi_{\|}(W)}{\partial x^{a}}+\frac{\partial\psi_{\bot}(W)}{\partial x^{a}}=\frac{\partial\psi(W)}{\partial x^{a}}$
then $\left[\tilde{ds},\tilde{\partial}_{s}\psi_{\|}\right]=0$ and
since $\psi(W)$ is a Hamilton-Jacobi function, it follows from Equation
\ref{eq:gr-eq-6a-1} and the definition of $\psi_{\bot}$ that 
\[
\tilde{ds}\,\tilde{\partial}_{s}\psi_{\|}(W)=d\psi_{\|}(W)\qquad\textrm{and}\qquad\left[\frac{\tilde{ds}}{ds},\frac{\tilde{\partial}\psi_{\bot}(W)}{\partial s}\right]
\]
are both integrable. Therefore, $\tilde{ds}\,\tilde{\partial}_{s}\psi(W)$
is an exact differential. The result follows.

\section{Clifford algebra and directional derivatives} 
\setcounter{section}{2}

Lemma \ref{lem:HJ-2-app} above is in its own way surprising. It requires
that the decomposition of a Hamilton-Jacobi non-linear wave function
can only be decomposed into a parallel component. In other words,
if $\psi(W)=\psi_{\|}(W)+\psi_{\bot}(W)$ where $[\frac{\tilde{ds}}{ds},{\frac{\tilde{\partial}\psi_{\|}(W)}{\partial s}}]=0$
then $\psi_{\bot}(W)=c_{o}t+c_{1}x_{1}+c_{2}x_{2}+c_{3}x_{3}$, with
$c_{o},c_{1},c_{2},c_{3}$ being constants. One might have expected that 
$\psi_{\bot}(W)$ would be arbitrary
and not simply linear. So why is this so? First of all, by definition
of $\psi_{\bot}(W)$ we expect that 
\[
\left\{\frac{\tilde{ds}}{ds},\frac{\tilde{\partial}\psi_{\bot}(W)}{\partial s}\right\}=0
\]
which means 
\[
\frac{d\psi_{\bot}(w)}{ds}=\frac{\partial\psi_{\bot}(W)}{\partial x^{i}}dx^{i}=0
\]
This implies that $\psi_{\bot}(W)=constant$ and this is the surprise.
In contrast, 
\[
\left[\frac{\tilde{ds}}{ds},\frac{\tilde{\partial}\psi_{\|}(W)}{\partial s}\right]=0
\]
does not imply $\psi_{\|}(W)=constant$. Usually in geometry, parallel
and perpendicular decompositions are relative to a basis. For example,
if ${\bf {x}=x_{1}{\bf {e}_{1}+x_{2}{\bf {e}_{2}}}}$ then the component
$x_{1}{\bf {e}_{1}}$ is parallel to ${\bf {e}_{1}}$ and perpendicular
to ${\bf {e}_{2}}$ while $x_{2}{\bf {e}_{2}}$ is perpendicular to
${\bf {e}_{1}}$ and parallel to ${\bf {e}_{2}}$ and vice-versa.
The choice of $x_{1}$ and $x_{2}$ are arbitrary and independent.
It might be instructive to take a concrete example and see what happens
with respect to the Clifford product. Consider 
\begin{eqnarray}
\tilde{ds}\tilde{\partial}_{s}\psi(W) & = & \left(\frac{\partial\psi}{\partial x}\gamma_{x}+\frac{\partial\psi}{\partial y}\gamma_{y}\right)(dx\gamma_{x}+dy\gamma_{y})\\
 & = & \frac{\partial\psi}{\partial x}dx+\frac{\partial\psi}{\partial y}dy+\gamma_{x}\gamma_{y}\left(\frac{\partial\psi}{\partial x}dy-\frac{\partial\psi}{\partial y}dx\right)
\end{eqnarray}
On the one hand, if $\left(\frac{\partial\psi}{\partial x},\frac{\partial\psi}{\partial y}\right)$
is parallel to $(\dot{x},\dot{y})$ then on letting $\frac{\partial\psi}{\partial x}=g(x,y)$
and $\frac{\partial\psi}{\partial y}=g(x,y)\frac{dy}{dx}$, we obtain
\[
\left[\frac{\tilde{ds}}{ds},\frac{\tilde{\partial}\psi_{\|}(W)}{\partial s}\right]=0
\]
and 
\begin{eqnarray}
\left\{ \frac{\tilde{ds}}{ds},\frac{\tilde{\partial}\psi_{\|}(W)}{\partial s}\right\}  & = & \left(\frac{\partial\psi}{\partial x}+\frac{\partial\psi}{\partial y}\frac{dy}{dx}\right)\dot{x}\\
 & = & \left[1+\left(\frac{dy}{dx}\right)^{2}\right]g(x,y)\dot{x}
\end{eqnarray}
On the other hand, if $\left(\frac{\partial\psi}{\partial x},\frac{\partial\psi}{\partial y}\right)$
is perpendicular to $(\dot{x},\dot{y})$ and we let $\frac{\partial\psi}{\partial x}=g(x,y)\frac{dy}{dx}$
and $\frac{\partial\psi}{\partial y}=-g(x,y)$, we obtain 
\[
\left\{ \frac{\tilde{ds}}{ds},\frac{\tilde{\partial}\psi_{\bot}(W)}{\partial s}\right\} =0
\]
and 
\begin{eqnarray}
\left[\frac{\tilde{ds}}{ds},\frac{\tilde{\partial}\psi_{\bot}(W)}{\partial s}\right] & = & \left(\frac{\partial\psi}{\partial x}\frac{dy}{dx}-\frac{\partial\psi}{\partial y}\right)\dot{x}\\
 & = & \left[\left(\frac{dy}{dx}\right)^{2}+1\right]g(x,y)\dot{x}
\end{eqnarray}
This demontrates the point that there is a one to one correspondence
between the set of projections onto the $\gamma_{x}$ and the $\gamma_{y}$
axes and yet when we imposed the property that $\psi$ should be an
\textbf{exact differential} this one to one correspondence breaks
down and we are forced to chose $g(x,y)=0$. This follows because
a total derivative, although as an inner product it is rotationally
invariant, nevertheless as a directional derivative defined relative
to a basis vector, its value varies according to the direction.


\begin{thebibliography}{99}
\bibitem{car-wp} E. Cartan, {\em The Theory Of Spinors} (Dover,1981),
p. 134. 

\bibitem{dn} G.Duff and D. Naylor, {\em Deifferential Equations of Applied Mathematics}, Wiley, New York, 1966, p 71.

\bibitem{jc-ca} J. Costella, B. McKellar, A. Rawlinson,{\em Am. J. Phys},\textbf{65},p. 835-841, 1997,

\bibitem{lh-re} L.Horwitz and R. Arshansky, {\em Phys. Lett. A}, \textbf{382}(26), p.1701-1708, 2018.

\bibitem{pm} P.Milonni, {\em The Quantum Vacuum}, Academic Press, New York, 1994, p. 322-323.

\bibitem{jo} John Oprea, {\em Differential Geometry and its Applications},
Pearson/Prentice Hall, New Jersey, 1997., p xv. 

\bibitem{ohara-wp} P. O'Hara, {\em Nuovo Cimento B}, \textbf{111}(7),
799-810(1996).

\bibitem{poh1-wp} P. O'Hara, {\em Foundations of Physics}\textbf{35}(9),
1563-1584(2005).

\bibitem{poh2-wp} P. O'Hara, {\em Jour. of Physics (Conf)},\textbf{330}(2011)012013.

\bibitem{poh-cm} P. O'Hara, {\em Jour. of Physics (Conf)},\textbf{437} (2013)012007. 

\bibitem{oraf-wp} L. O'Raifeartaigh, {\em The Dawning of Gauge
Theory} (Princeton University Press, 1997), p 112. 

\bibitem{schr} E. Schrodinger, {\em Ann. d. Physik}, {\textbf [81]} (1926).

\bibitem{sg} J. Synge and B. Griffith,{\em Principles of Mechanics},
McGraw-Hill, 1959,pp.440-463.
\end{thebibliography}
\end{document}